# Numerical modeling of hydrogel scaffold anisotropy during extrusion-based 3D printing for tissue engineering

**V.T. MAI[a,b], R. CHATELIN[c], E.-J. COURTIAL[d], C. BOULOCHER[e,f], R. RIEGER[a,b]**

a. Université de Lyon, VetAgro Sup, UPSP ICE 2021.A104, France
b. Université de Lyon, École Centrale de Lyon, France
c. Université de Lyon, École Centrale de Lyon–ENISE, LTDS UMR CNRS 5513, France
d. 3d.FAB, Université de Lyon, Université Lyon 1, CNRS, INSA, CPE-Lyon, ICBMS, Villeurbanne Cedex, France
e. UniLaSalle Polytechnic Institute, Veterinary College, Campus of Rouen, 76130 Mon Saint Aignan, France
f. Transformations et Agroressources, ULR 7519, Institut Polytechnique UniLaSalle, Université d'Artois, 76130 Mon Saint Aignan, France

Corresponding author: romain.rieger@ec-lyon.fr (R. Rieger)

Authors:
van.mai@vetagro-sup.fr (V. T. Mai)
robin.chatelin@enise.fr (R. Chatelin)
edwin.courtial@univ-lyon1.fr (E.-J. Courtial)
caroline.boulocher@unilasalle.fr (C. Boulocher)
romain.rieger@ec-lyon.fr (R. Rieger)

## Abstract

Extrusion-based 3D printing is a widely utilized tool in tissue engineering, offering precise 3D control of bioinks to construct organ-sized biomaterial objects with hierarchically organized cellularized scaffolds. The internal organization of scaffold constituents must replicate the structural anisotropy of the targeted tissue to effectively promote cellular behavior during 3D cell culture. The choice of polymers in the bioink and extrusion process topological properties significantly impact tissue engineering constructs' structural anisotropy and cellular response. Our study employed a hydrogel bioink consisting of fibrinogen, alginate, and gelatin, providing biocompatibility, printability, and shape retention post-printing. Topological properties in flowing polymers are determined by macromolecule conformation, namely orientation and stretch degree. We utilized the micro-macro approach to describe hydrogel macromolecule orientation during extrusion, offering a two-scale fluid behavior description. The study aimed to use the Fokker-Planck equation to represent constituent population (polymer chain) state within a hydrogel's representative elementary volume during extrusion-based 3D printing. Our findings indicate that a high shear rate drives constituent orientation in tubular nozzle syringe setups, overcoming fluid rheological behavior. Additionally, the interaction coefficient ($C_i$), representing microscopic fluid particle interaction, surpasses hydrogel behavior for constituent orientation prediction. This approach provides an initial but robust framework to model scaffold anisotropy, enabling optimization of the extrusion process while maintaining computational feasibility.

# 1. Introduction

Tissue engineering (TE) is a set of techniques inspired by the principles of engineering and life sciences, which is used to develop biological substitutes capable of restoring or maintaining the function of specific tissues [1]. The objective of TE is to recreate patient-specific organs or tissues on demand to restore physiological functions. In addition, engineered tissue can be used to test new drugs or cosmetics without resorting to animal testing. Among the available techniques, extrusion-based 3D printing is a recent and widely used TE tool that provides precise 3D control of bioinks in order to create organ-size biomaterial-based objects [2,3]. With this technology, bioinks such as hydrogels can be used to create hierarchically organized cellularized scaffolds *in vitro*. Extrusion forces such as compressed air or a mechanical screw extrude cellularized or acellularized hydrogel filaments, layer-by-layer, from a cartridge through a nozzle or needle [4]. Extrusion-based 3D printing is the most commonly employed technique in bioprinting due to its great potential and compatibility with different cell types and densities, and large range of viscosity, from $3 \times 10^{-2}$ to $6 \times 10^4$ Pa.s [5–7].

The rheological behavior of hydrogels prior to extrusion and the printing conditions (extrusion parameters, nozzle dimensions) have a significant impact on printability *i.e.* extrudability and shape fidelity of the scaffold [8,9]. Once the hydrogel has been printed, crosslinking mechanisms are activated to fix the shape of the scaffold [4,5], for example, via pH [10], photochemical reactions [11], enzymatic cross-linking [12] or temperature [13]. These processes ensure that the printed scaffold retains its intended shape. Cellularized scaffolds are then grown in a bioreactor to achieve TE construct maturation by promoting cell-extracellular matrix interaction and activation of mechano-sensitive molecular pathways in cells. During this phase, scaffold mechanical properties and topological features, especially the orientation of scaffold constituents, play a crucial role in the stimulation of different cellular behaviors, such as migration, proliferation, differentiation and extracellular matrix synthesis [14–16]. Furthermore, the orientation of scaffold constituents affects cellular morphology [17–19]. Therefore, internal organization of scaffold constituents to mimic the structural anisotropy of targeted tissue and stimulate proliferative cellular behavior is of primary concern during 3D cell culture for TE [20,21].

The selection of polymers, also known as fibrillar proteins, that make up the hydrogel, along with the topological properties during the extrusion process, significantly impact the cellular response in fibrillar hydrogels [8,22]. These factors contribute to the formation of anisotropic fibrillar networks that closely resemble those found in native tissues, resulting in distinct mechanical properties and mechanobiological behaviors, as opposed to non-fibrous inks which often result in random cell organization and isotropic fibrillar networks [23]. The hydrogel used in our study was made of three constituents: fibrinogen, alginate and gelatin (FAG). These components provide biocompatibility (an environment favorable to cell survival), printability (appropriate extrusion quality) and conservation of the 3D shape after printing [24–26]. This study represents a first approach to understanding flow-induced anisotropy in bioprinting to provide a predictive framework for scaffold design.

The topological properties in flowing polymers are dictated by macromolecule conformation, such as their orientation and degree of stretch. However, achieving chain orientation is challenging, as it requires temperatures close to the glass transition temperature (Tg), induced by forces that promote deformation in the printing direction. These forces are lower and more achievable near Tg, as opposed to below Tg, where they need to be much higher due to the viscoelastic solid-like behavior of the material. Conversely, when operating close to or above Tg, the fluid-like behavior of the material leads to very rapid relaxation times, causing the polymer chains to revert to a randomly oriented state [27]. Ghodbane et al. [27] demonstrated that aligning polymer chains is possible by moving the printhead at sufficiently high speeds while the polymer remains in a semi-solid state as it cools from the fluid state at the nozzle tip. This alignment was achieved through a precise combination of nozzle diameter, extrusion pressure, and temperature. Their findings revealed that the level of orientation obtained, as measured by X-ray diffraction and thermal shrinkage, was superior to that of conventionally drawn fibers.

Anisotropic fibrillar hydrogels induced by shear-flow have been investigated using various 3D printing techniques, including laser-based methods such as stereolithography [28–31] and extrusion-based 3D printing [32–35]. These studies consistently report the formation of anisotropic extracellular matrix structures, driven by shear-induced alignment of fibrillar networks during the printing process. Although some research has estimated the shear levels using analytical calculations [32,34] or numerical methods such as finite element analysis [28,30], none have quantified the degree of orientation distribution of the hydrogel constituents during 3D printing, which is crucial for achieving fibrillar anisotropy in scaffolds. Additionally, there is currently no quantitative data linking the orientation of these constituents to specific shear levels based on the rheological behavior of a given ink.

The study of hydrogel scaffold anisotropy during extrusion-based 3D printing requires balancing model complexity and computational feasibility. In this work, we employ a one-way coupling approach, where polymer orientation responds to the flow field without influencing it. Inspired by Folgar and Tucker's extension of Jeffery's equation, the introduction of a diffusion term into Jeffery's equation allows us to account for particle interactions through the interaction coefficient $C_i$ in semi-dilute regimes without requiring a fully coupled simulation. This methodology aligns with previous studies [36,37] that have shown that one-way coupling provides comparable results to fully coupled models for fiber orientation under most flow conditions. To achieve this, we used the micro-macro approach to describe the orientation state of hydrogel macromolecules during the extrusion process, offering a two-scale description of the fluid behavior. The goal of our study was to use the Fokker-Planck equation, which describes the evolution of the probability distribution function (PDF) over time, to represent the real state of the constituent population (polymer chains) in the representative elementary volume (REV) within a hydrogel during extrusion-based 3D printing. This is similar to previous work modeling the fiber-reinforced composite injection molding process [38,39]. While these simplifications do not capture the complete feedback mechanisms, especially those related to polymer-fluid interactions at higher concentrations, they provide a robust foundation for modeling hydrogel anisotropy with reasonable computational effort, particularly for applications in tissue engineering. To the authors' knowledge, this is the first published report of micro-macro modeling using the PDF approach to describe the orientation state of hydrogel constituents during extrusion-based 3D printing.

By establishing a direct link between shear levels and the orientation of hydrogel constituents, our study enables the quantification and optimization of printing parameters to achieve controlled anisotropy in scaffolds. This methodology, although demonstrated on a specific hydrogel formulation, can be readily adapted to other bioink compositions within the assumptions and constraints of the current model, as it only requires knowledge of the ink's rheological behavior. While our work primarily focuses on the extrusion process, the same principles can be applied to other 3D printing techniques, such as stereolithography. In such cases, the local orientation of constituents can be quantified and fine-tuned using photopolymerization, providing precise control over the anisotropic properties of the printed constructs.

While our model provides valuable insights into the anisotropy of hydrogels, there are inherent limitations that should be addressed. Future experimental validation, particularly through techniques like X-ray synchrotron analysis, is essential to confirm our predictions and further refine the model. Such experimental data could also help overcome some of the assumptions made in this first attempt and pave the way for more advanced and realistic simulations. Prediction of the orientation of hydrogel constituents during the extrusion process could give new insights into the structural anisotropy of scaffolds, allowing tissue engineers to achieve convenient cellular behavior and specific mechanical properties.

## 2. Materials and methods

### 2.1 Rheological characterization of the hydrogel formulation

The ink used in this study was a hydrogel made of three main components: 2% (w/v) fibrinogen, 4% (w/v) alginate and 2% (w/v) gelatin, as previously described [24]. Before rheological measurement, the hydrogel was stored for 30 min at 21°C to meet printing conditions and stabilize the rheological properties [3]. A DHR2 rotational rheometer (TA instruments, Texas, USA) with a concentric cylinder geometry was used to characterize the rheological behavior of the hydrogel. The analysis was carried out in triplicate based on the shear rate scale [0.01, 100] $s^{-1}$ at 21°C. Finally, the curves of shear stress versus shear rate were plotted showing the flow diagrams of the hydrogel. From the basic rheological data curves, a mathematical model describing hydrogel behavior and rheological parameters was determined by a numerical least square minimization using a Levenberg-Marquardt nonlinear regression algorithm. It is important to note that the hydrogel used for these experimental tests did not contain any cells to allow an initial investigation of constituent orientation during extrusion.

### 2.2 Flow modeling

Flow modeling was performed using Comsol Multiphysics (version 5.6, COMSOL, Inc.). The finite element method (FEM) was used to solve the fluid flow problem, using a classical second order finite element for the velocity and a first order element for the pressure to ensure computational compatibility. For the hydrogel used in this study, the fluid flow governing equations are given by Navier-Stokes equations [40]. As our work focused on a stationary solution, these equations are given by:

$$\begin{aligned}
\nabla.\vec{u} &= 0 \\
\rho\nabla\vec{u}.\vec{u} &= -\nabla q + \nabla.[\mu(\nabla\vec{u} + \nabla\vec{u}^T)]
\end{aligned} \tag{1}$$

The first equation ensures fluid incompressibility, while the second guarantees momentum conservation. Hydrogel density is assumed to be constant, so the mass conservation equation is not solved. In the equations (1), $\vec{u}$ denotes the fluid velocity, $q$ the pressure of the fluid $[Pa]$, and $\rho$ and $\mu$ are the density $[kg/m^3]$ and dynamic viscosity $[Pa.s]$ of the hydrogel, respectively.

The associated Reynolds number ($Re$) was calculated to characterize the flow regime as follows:

$$Re = \frac{\rho U l}{\mu} \tag{2}$$

where $U$ is the average velocity $[m/s]$ and $l$ is the characteristic length $[m]$.

Further flow characterization was conducted to account for the non-Newtonian behavior of the hydrogel, specifically for the Herschel-Bulkley fluid model. In this case, the Bingham number ($Bm$), was calculated to characterize the relationship between yield stress and viscous stress:

$$Bm = \frac{\tau_0 l}{\mu U} \tag{3}$$

where $\tau_0$ is the static yield stress $[Pa]$, $l$ the characteristic length $[m]$, $U$ the average velocity $[m/s]$, and $\mu$ the dynamic viscosity $[Pa.s]$. Additionally, for numerical stabilization in the context of convection-dominated regimes, the Péclet number ($Pe$) was computed using the artificial diffusion coefficient:

$$Pe = \frac{U l}{D_{num}} \tag{4}$$

where $D_{num}$ is the artificial diffusion coefficient $[m^2/s]$. In this study, we set $D_{num} = 10^{-5}$ [36], the Péclet number was computed accordingly.

## 2.3 Orientation of hydrogel constituents

The motion of a single particle in a dilute solution was first described by Jeffery *et al.* [41], who assumed the particle to be a rigid ellipsoid immersed in a viscous Newtonian fluid. However, this prediction model is valid only for dilute suspensions and does not consider the interaction between particles, which has a significant impact on their orientation within hydrogels. Orientation tensors are another approach widely used in the literature to describe the orientation state in mechanical behavior modeling [42] because the tensor description simplifies the computation. A major drawback of orientation tensors is the requirement of the so-called closure equation to approximate the higher order tensors that occur in the orientation evolution equations. This may lead to the loss of physical description and accuracy. The Hermans coefficient can be used to calculate the average orientation of macromolecule chains in polymer solution [43]. This approach requires the choice of a reference axis, and the orientation of the macromolecule is then determined by its angle to this axis. Another method to predict the orientation of the constituents in a REV is to use the PDF as a statistical approach. One big advantage of the PDF approach is the computational accuracy and the possibility to define the relationship between initial and final orientations. Moreover, the PDF can represent polymer chain conformations at specific points [44–46].

The numerical simulation presented in this article investigates constituent orientation in a hydrogel (a polymeric fluid) during bioprinting. Therefore, the modeling includes both fluid and constituents. The rigid dumbbell model is the simplest model for representing the configuration of a polymer chain. Using this model, the orientation of a constituent can be represented by a unit vector $\vec{p}$ aligned along the principal particle axis. The direction and coordinates of the vector $\vec{p}$ with respect to the vertical axis (y-axis) in a 2D Cartesian system are shown in Figure 1.

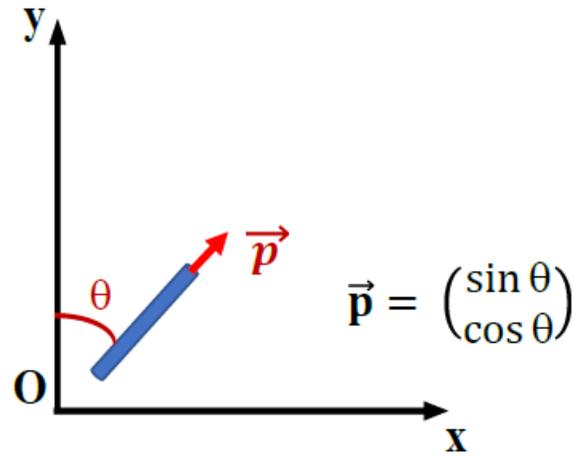

Figure 1. Definition of the orientation angle vector $\vec{p}$ in a 2D Cartesian system.

In industrial applications, it is difficult to follow the movement of each individual particle in a suspension due to the computational cost. Therefore, REV can be used to derive the averaged macroscopic quantities of many constituents. The approach used in this study employs a statistical point of view to describe constituent orientation. Accordingly, we used a PDF to develop a numerical model to predict the flow-induced constituent orientation in a REV. The PDF $\psi(\vec{x}, \vec{p}, t)$ defines the probability of finding at time $t$ a constituent $\vec{p}$ oriented according to an angle $\theta$ with respect to the vertical axis (y-axis) at a point $\vec{x}(x, y)$ defined by Eulerian coordinates in the 2D Cartesian system (see Fig. 1). Thus, the function $\psi$ must satisfy certain physical conditions [42]. Firstly, a constituent oriented at any angle $\theta$ is indistinguishable from a constituent oriented at angle $\pi - \theta$, so $\psi$ must be periodic leading to:

$$\psi(\vec{p}) = \psi(-\vec{p}) \tag{3}$$

Secondly, as $\psi$ is a PDF, it must satisfy the normalization condition:

$$\int \psi(\vec{p}) d(\vec{p}) = 1 \qquad (4)$$

The final condition, related to the continuity condition, describes the change of $\psi$ with respect to time when the constituents are changing and can be expressed as:

$$\frac{\partial \psi}{\partial t} + \vec{u}. \nabla_x \psi = -\frac{\partial}{\partial \vec{p}}. (\psi \dot{\vec{p}}) \qquad (7)$$

where $\vec{u}$ is the fluid velocity. The constituent rate of change $\dot{\vec{p}}$ is given by the Jeffery equation [41]:

$$\dot{\vec{p}} = \mathbf{W}. \vec{p} + \lambda [\mathbf{D}. \vec{p} - \mathbf{D} : (\vec{p} \otimes \vec{p}) \vec{p}] \qquad (8)$$

where $\mathbf{W} = \frac{1}{2} [(\nabla \vec{u}) - (\nabla \vec{u})^T]$ is the vorticity tensor and $\mathbf{D} = \frac{1}{2} [(\nabla \vec{u}) + (\nabla \vec{u})^T]$ is the deformation rate tensor. The coefficient $\lambda = (L^2 - D^2)/(L^2 + D^2)$ is a scalar particle shape depending on the aspect ratio of the constituent with length $L$ and diameter $D$. The Folgar and Tucker diffusion term $D_r \nabla_p^2 \psi$ is then introduced into the Jeffery equation (8) in order to simulate particle interactions in the semi-concentrated or concentrated regime [47]. The continuity equation (7) finally becomes:

$$\frac{\partial \psi}{\partial t} + \vec{u}. \nabla_x \psi = -\frac{\partial}{\partial \vec{p}}. (\psi \dot{\vec{p}}) + D_r \nabla_p^2 \psi \qquad (9)$$

where $\nabla_p^2$ refers to $\partial^2 / \partial \vec{p}^2$, $D_r = C_i |\dot{\gamma}|$, with $|\dot{\gamma}| = \sqrt{2 \mathbf{D} : \mathbf{D}}$ denoting the scalar magnitude of the deformation rate tensor and $C_i$ the interaction coefficient. Equation (9), well known as the Fokker-Planck equation, is the most general, complete and unambiguous description of the constituent orientation evolution. The numerical strategy to solve the Fokker-Planck equation is detailed in the next section to estimate a numerical prediction of flow-induced constituent orientation.

## 2.4 Numerical strategy of the micro-macro approach

Several approaches have been used to model the orientation state of macromolecules [48]. The first uses atomistic flow simulations, which is currently restricted to very coarse models of polymers and molecular size flow geometries. The second uses the kinetic theory to describe the evolution of a coarse-grained model of the polymer macromolecule on a microscopic scale [49–51]. In combination with macroscopic conservation laws, this computational rheology procedure constitutes the so-called micro-macro approach. The third approach is the macroscopic approach using continuum mechanics [52,53].

When dealing with anisotropic flowing polymers, the micro-macro approach involves an averaging procedure because it is not possible to follow every macromolecule orientation. There are two different methods of micro-macro description [38]. The first is spatial averaging or homogenization, enabling the derivation of macroscopic quantities as soon as the microscopic orientation is specified. The second is statistical averaging as the orientation cannot be precisely described. This method enables researchers to extract averaged macroscopic quantities from a statistical description of the orientation. The simplest description of macromolecule orientation using kinetic theory is the dumbbell model, which considers two beads connected by a rod (bead-rod model) or a spring (bead-spring model), and can behave as linear (Hookean model) or non-linear (Finitely Extensible Non-linear Elastic (FENE) model) [49,51,54]. Bead-rod models can accurately describe the rheological behavior of dilute polymer solutions [55,56]. Macromolecules comprising biological structures, such as polypeptides, fibrous proteins (*e.g.* keratin, fibrinogen, collagen and actin), DNA and some viruses (such as tobacco mosaic virus), can be considered to be rigid rods [57,58].

The rigid rod (or dumbbell) model used here offers a simplified representation of polymer behavior, enabling us to focus on the primary effect of shear-induced alignment. Although this model neglects polymer stretching and entanglement, which become important at higher concentrations, it allows for manageable computational demands. Doi and Edwards [50], have demonstrated that these models provide valuable insights into the relationship between macromolecular motion and rheological phenomena, especially in dilute and semi-dilute regimes. This simplification is necessary for this initial study, but future models will aim to include more complex polymer dynamics.

In this section, a numerical approach is generalized to compute the evolution of constituent orientation governed by the Fokker-Planck equation. Our study was restricted to a steady flow in 2D, therefore the equation (9) can be rewritten as:

$$\vec{u}.\nabla_x \psi = -\frac{\partial}{\partial \vec{p}}.\left(\psi \dot{\vec{p}}\right) + D_r \nabla_p^2 \psi \qquad (10)$$

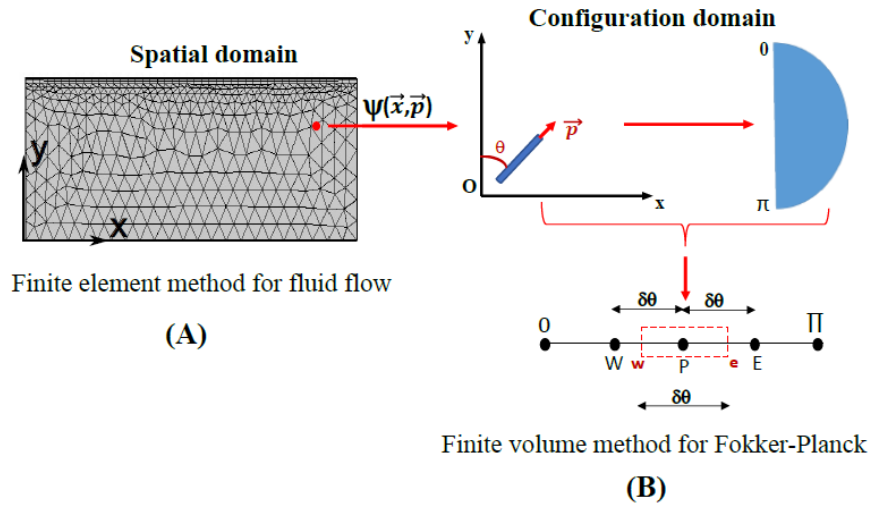

Figure 2. Schematic diagram representing the procedure for the numerical resolution of the Fokker-Planck equation. (A) The 2D fluid flow governed by the Navier-Stokes equations is solved using a finite element discretization in the spatial domain. It is coupled to (B), a finite volume discretization in the configuration domain. P, E and W represent the central, east and west nodes of the finite volume system, respectively. e and w indicate east and west edges of the control volume, respectively.

Figure 2 depicts a schematic diagram in which the Fokker-Planck equation will be solved in order to determine the PDF ($\psi$) at each point on the flow grid considered as the constituent position $\vec{x}(x, y)$ in Cartesian coordinates defining the spatial domain, and the angle $\theta$ between the constituent axis and the y-axis defining the configuration domain. Consequently, the difficulty in solving the Fokker-Planck equation lies in the fact that it must be concurrently solved in the spatial domain for the fluid flow and the configuration domain for the orientation. Although there are only two space variables, it is still a three-dimensional problem because there is an extra angle variable as presented in Figure 2.

Therefore, we used Comsol Multiphysics (v5.6) to implement a dedicated numerical strategy with two discretization steps [36,59,60]. The first discretization uses FEM as depicted in Figure 2A to solve the fluid problem. FEM results then provide the velocity gradient at each point of the grid, which is used in the Fokker-Planck equation to evaluate the constituent orientation. (ii) The second discretization uses the Finite Volume Method (FVM) as shown in Figure 2B to solve the Fokker-Planck equation (10) for PDF. In 2D, all possible orientations defined by the angle $\theta$ describe a half circle of unit radius, as shown in Figure 2B. At this point, the general central node is identified by $\vec{p}$ in the control volume and its neighbors W and E (west and east, respectively) in a one-dimensional geometry. The key step of this Finite Volume discretization is the integration of the Fokker-Planck equation over this control volume to establish a system of linear algebraic equation, which yields the following form:

$$\delta\theta. \vec{u}. \nabla_x \psi_P + a_P \psi_P - a_W \psi_W - a_E \psi_E = 0 \qquad (11)$$

To evaluate the Fokker-Planck equation at the control volume edges, an upwind differencing and power law scheme was employed to ensure the transportiveness and conservation volume of the PDF for the whole solution domain [61]. Therefore, the coefficients $a_P, a_W, a_E$ were determined from these approaches and are presented in Appendix A. $\delta\theta$ represents the control volume width. In this study, half of a circle $[0, \pi]$ was divided into 30 nodes corresponding to 30 variables for the PDF [33], thus $\delta\theta = \pi/30$. At this stage, a system of 30 analytic equations of equation (11) was implemented into the partial differential equations interface of Comsol Multiphysics as a convection-diffusion problem. This strategy enabled the resolution of the configuration domain as a FVM discretized system of equations coupled to the resolution of the spatial domain as a FEM discretized fluid problem.

In this study, the following assumptions were made. First, constituents were considered to be rigid particles, uniform in length and diameter. Second, the constituent aspect ratio was set to $\lambda = 1$ to avoid numerical difficulties associated with the tumbling effect in a flow. Also, setting $\lambda = 1$ allowed us to obtain an equation for the motion of small, rigid dumbbell-like particles, such as hydrogel constituents [42,62,63]. The interaction coefficient $C_i$ in Folgar and Tucker's term was first fixed to $C_i = 0.01$ to validate the model, then a parametric study on $C_i$ was performed in the range of $C_i \in [0.0033; 0.1]$ to study its effect on the constituent orientation prediction model [47].

The PDF results indicate the probability of constituent orientation at every grid point of the flow. Thus, constituent orientation can be represented by orientation ellipses calculated directly from the

PDF functions obtained after simulation through the so-called second order orientation tensor **A** as:

$$A := \int \vec{p} \otimes \vec{p}\,\psi(\vec{p})\,d\vec{p} \tag{12}$$

The eigenvectors calculated from the second order orientation tensor **A** give the length and direction of the two major axes of the ellipse, and the eigenvalues give the alignment magnitude of constituent orientation along these directions. Therefore, an orientation ellipse can be generated as a graphical representation of the average constituent orientation at each point in the grid. All the post-processing steps were performed using Comsol LiveLink for MATLAB (v5.6 and R2018a).

## 2.5 Computational geometries and boundary conditions

This work focused on the constituent orientation in a 2D flow. The first geometry considered was a planar channel as an academic test-case to investigate both simulation performances and the qualitative effect of the parameters. The study of constituents in a channel domain was used to validate the model. Then, as a practical application, we performed an extension through more complex geometry as a 2D model of a syringe with a tubular nozzle dedicated to 3D bioprinting.

The geometries and boundary conditions used are shown in Figure 3A for a planar channel, which serves as a benchmark, and Figure 3B for a tubular nozzle syringe. The planar channel consisted of a 4-mm-wide and 2-mm-high rectangle. The boundary conditions (BC) were as follows: the inlet **BC1** at $x = 0$ mm was set as $380\ Pa$ pressure imposed, the outlet **BC2** at $x = 4$ mm was set to $0\,Pa$. The syringe was a tubular nozzle type, 30 mm in length, with an inner diameter of 200 µm (Optimum®, Nordson EFD, USA). The boundary conditions were as follows: a constant flow rate of $5\mu l/s$ was applied at the inlet **BC1'** in order to generate the extrusion process, as this value is compatible with bioprinting conditions in many applications [3]. At the nozzle exit **BC2'**, a standard atmospheric pressure was imposed as a real extrusion condition set to 1 atm = 1.01325 bar. For both geometries, a no-slip wall condition **BC3** and **BC3'** was applied on the top meaning $\vec{u} = 0$ along the wall.

Axial symmetry can be used to reduce computational cost and keep physical accuracy so that only half of the domain is considered. Therefore, **BC4** and **BC4'** are symmetry conditions defined as $\vec{u}.\vec{n} = 0$ where $\vec{n}$ is the surface normal vector. Both geometries are discretized by using the default mesh which consists of 3323 triangles for the channel and 19989 triangles for the syringe. Finally, for the Fokker-Planck equation, the constituents are assumed to be isotropic leading boundary conditions **BC1** and **BC1'** for the channel and the syringe set to $\psi = 1/\pi$, respectively.

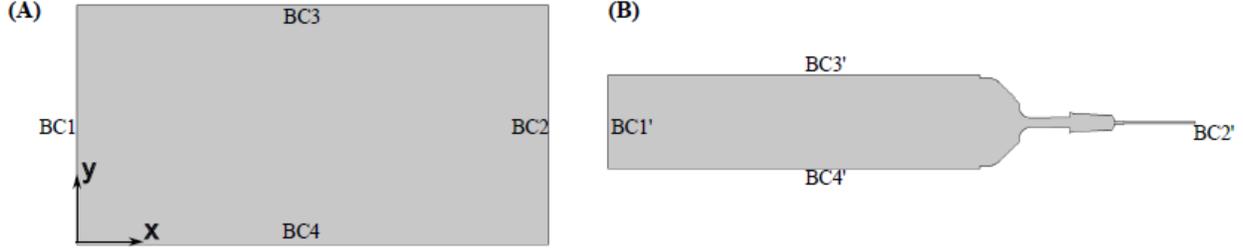

Figure 3. 2D geometries and boundary conditions: (A) Benchmark planar channel - **BC1**: $P_{in} = 380$ Pa and $\psi = 1/\pi$; **BC2**: $P_{out} = 1.01325$ Pa ; **BC3**: no-slip ; **BC4**: symmetry condition. (B) Tubular nozzle syringe - **BC1'**: $V_{in} = 5\ \mu l/s$ and $\psi = 1/\pi$ ; **BC2'**: $P_{out} = 1.01325$ Pa ; **BC3'**: zero-slip ; **BC4'**: symmetry condition.

We considered both Newtonian and Herschel-Bulkley hydrogel rheological behaviors. We used the Herschel-Bulkley fluid values of a typical hydrogel with a density $\rho = 1035$ [kg/m$^3$] and determined the other parameters of the model from the rheological measurement presented in the next section. The incompressible Newtonian fluid was employed with a density of 850 [kg/m$^3$] and a dynamic viscosity of 0.8 [Pa.s] for the channel and 10 [Pa.s] for the syringe. These parameters were adjusted for the Herschel-Bulkley fluid behavior for the constituent orientation analysis within a similar field velocity in order to get fair comparisons between Newtonian and non-Newtonian computations.

# 3 Results and discussion

## 3.1 Rheological characterization

The rheological diagram of the FAG ink is shown in Figure 4.

The shape of the experimental data indicates that the rheological behavior of the FAG can be modeled by the Herschel Bulkley law, which considers both the yield stress and shear thinning. The rheological equation of this model is given by the following relation:

$$\tau = \tau_0 + k\dot{\gamma}^n \tag{13}$$

where $\tau$ is the shear stress [Pa], $\tau_0$ the static yield stress [Pa] $k$ the consistency index [Pa.s], $\dot{\gamma}$ the shear rate [s] and $n$ the flow behavior index [-]. A home-made MATLAB program using the Levenberg-Marquardt nonlinear regression algorithm enabled us to obtain the Herschel-Bulkley model parameters: $\tau_0 = 113.96$ [Pa], $k = 195.36$ [Pa.s] and $n = 0.43$.

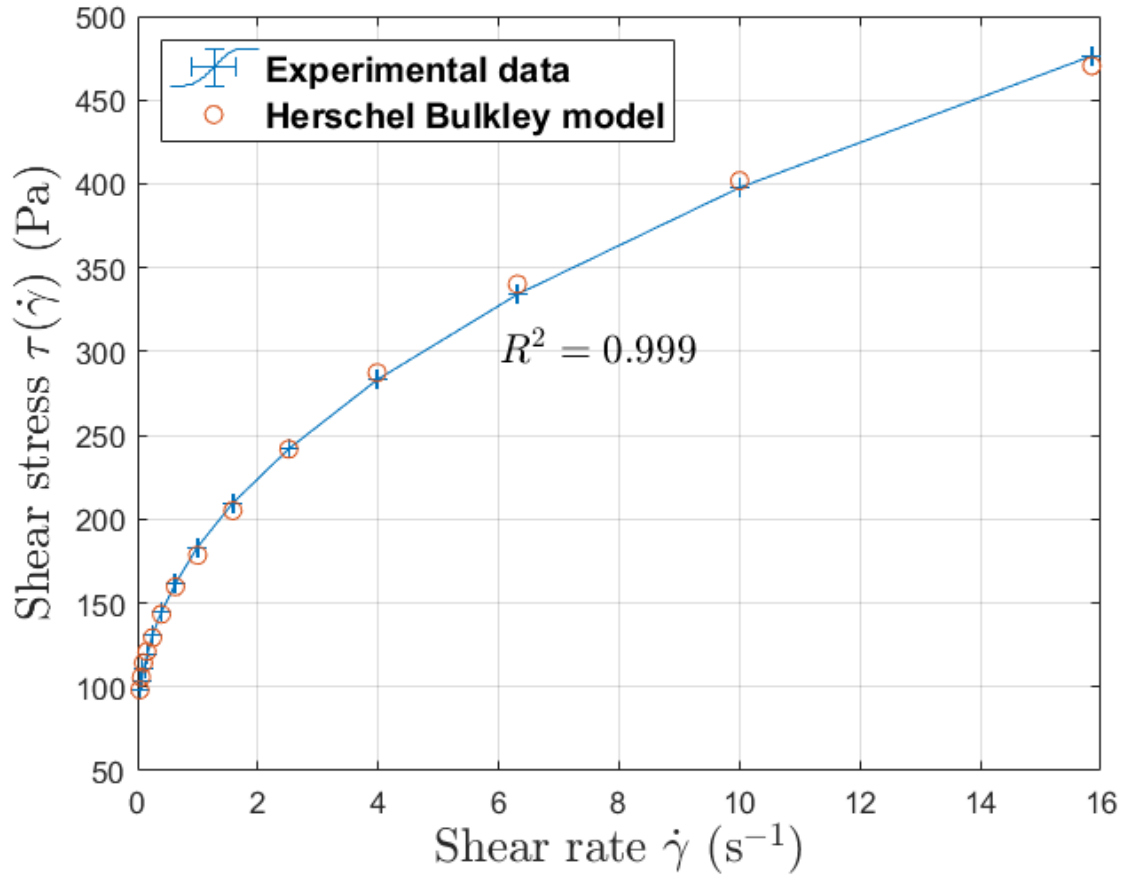

Figure 4. Rheological diagram of the FAG hydrogel. The error bars in the plots represent the relative error between the experimental data and the Herschel-Bulkley model. The coefficient of determination (R²) is an indicator that assesses the goodness of fit of a model.

Figure 4 illustrates the good numerical fit of the Herschel Bulkley model with experimental data. The results show a constant non-zero shear function $\tau(\dot{\gamma})$ at the origin ($\dot{\gamma} = 0$), indicating the static yield stress $\tau_0$ as the limit of the linear viscoelastic range at which the hydrogel starts to flow. Below this value, the material retains a solid-like behavior that opposes the flow. The model allows prediction of the shear stress required to initiate extrusion of the hydrogel. When the shear stress increases above the yield stress, the material rapidly breaks down and the hydrogel behaves as a liquid. Also, with its shear thinning behavior, the material is liquefied due to the increasing shear rate when passing through a smaller section. Therefore, the Herschel-Bulkley behavior, which takes into account the yield stress and shear thinning phenomenon, is the most used rheological behavior to assess printability in 3D printing processes [4,64].

## 3.2 Flow modeling

For targeted bioprinting applications, the Reynolds number computation leads to: $Re = 0.51$ for Newtonian fluid and $Re = 2.54 \times 10^{-3}$ for Herschel-Bulkley fluid in a channel geometry, which serves as a benchmark; $Re = 4.08 \times 10^{-3}$ for Newtonian fluid and $Re = 2.12 \times 10^{-4}$ for Herschel-Bulkley fluid in a tubular nozzle syringe geometry (see section 2.5 for a description of geometries). These results show that the Re numbers are compatible with a viscous fluid with a laminar flow. Additionally, the Bingham number ($Bm$) was computed for the Herschel-Bulkley fluid model, yielding values in the range of [0.00077 - 0.62]. The small values of $Bm$ indicate the dominance of viscous stresses over yield stresses during the extrusion process. Furthermore, the Péclet number ($Pe$) was found in the range of [9.6 to 120], which is consistent with convection-dominant flow conditions in the extrusion process, confirming the need for numerical stabilization via an artificial diffusion coefficient.

### 3.3 Hydrogel constituents orientation during extrusion-based 3D printing for TE scaffolds

#### 3.3.1 Poiseuille flow through planar channel as a benchmark

We used a classical benchmark test—a suspension hydrogel flow passing through a planar channel [36,65,66]— to investigate the validity of the constituent orientation prediction model. It should be noted that, in this work, a Newtonian fluid was employed to simplify the analysis of hydrogel flow. Although Newtonian behavior does not represent a realistic case in the context of 3D bioprinting, utilizing the Newtonian flow model allows for the rapid generation of results with less expensive computation times, making it a valuable choice for establishing a benchmark. The background color in Figure 5A and 5B shows the velocity profile in the xy-plane domain for both Newtonian and Herschel-Bulkley behaviors. For the Newtonian behavior, a fully developed Poiseuille flow was observed for which the parabolic velocity gradient increased linearly from the axisymmetric axis ($y = 0$) to the wall ($y = 2$) where a no-slip boundary condition (BC2) was defined. For the Hershel-Bulkley behavior, the classical plug velocity profile was obtained.

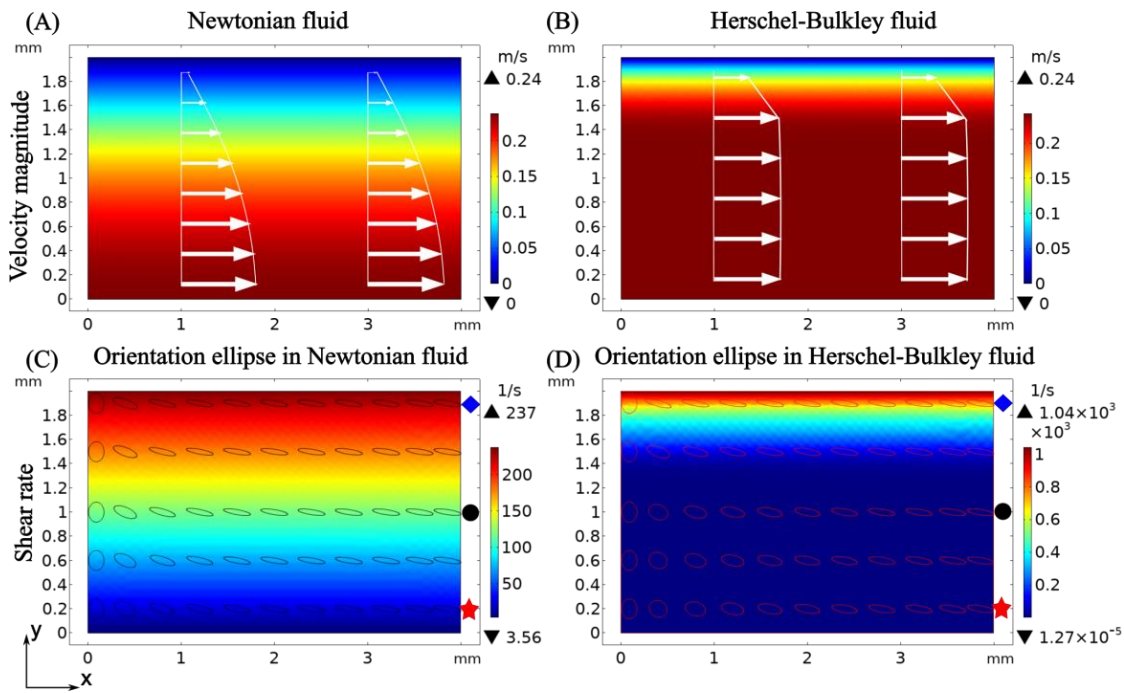

Figure 5. Velocity magnitude and shear rate distribution in the computational domain for the benchmark with interaction coefficient $C_i = 0.01$. The velocity profile is shown by white arrows,

the average constituent orientations are represented by the black and magenta ellipses. (A) and (C) Newtonian behavior; (B) and (D) Hershel-Bulkley behavior.

The shear rate distribution is shown in Figure 5C and 5D, with the ellipses in the foreground representing the average constituent orientations. The numerical results shown in Figure 5C and 5D highlight the preferred orientation of constituents at the center of each ellipse in the computational domain. Circle-like ellipses illustrate isotropic constituent orientation, which corresponds to the BC1 on the left-hand side of the model applied to the Fokker-Planck equation such that $\psi = 1/\pi$. Ellipses are then flattened and elongated, showing a preferential alignment indicated by their major axis orientation. They gradually align along the channel exit (x-axis direction) toward the flow direction but not completely, as illustrated in Figure 6 by the peak at $\theta \approx 1.7$. The anisotropy is such that $\theta \neq \pi/2$ due to the interaction between the constituents modeled by the interaction coefficient $C_i = 0.01$, which is non zero. Furthermore, along the y-axis, the constituents reach a steady state orientation more quickly along the x-axis direction near the wall ($y = 2$) than near the axisymmetric axis ($y = 0$), where the steady state is reached further downstream.

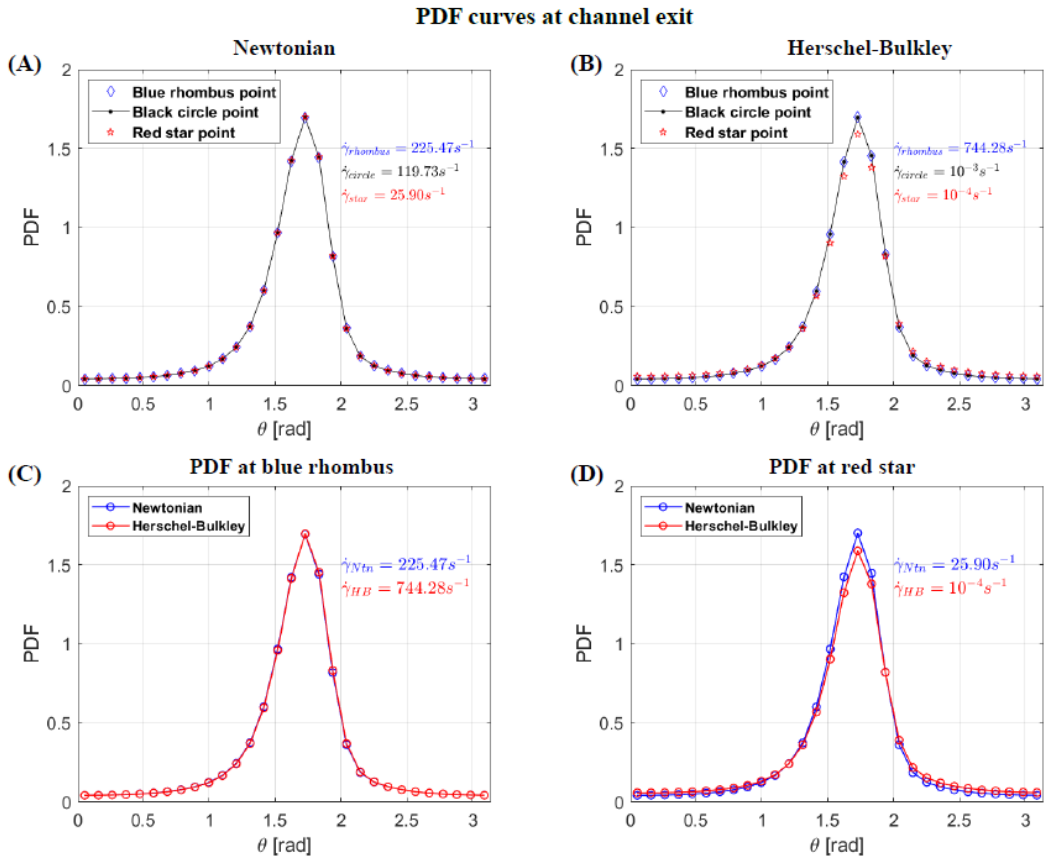

Figure 6. PDF curves for the benchmark with interaction coefficient $C_i = 0.01$ showing constituent orientation at three points (blue rhombus, black circle and red star) in Figure 5. (A) Newtonian behavior; (B) Herschel-Bulkley behavior; (C) and (D) PDF curves compared between Newtonian and Hershel-Bulkley behavior at blue rhombus and red star points, respectively.

We then analyzed the influence of the shear rate on the PDF distribution. Figure 6A and 6B illustrate the PDF distribution from higher to lower shear rate (the blue rhombus and black circle points on Figure 5C and 5D show higher shear rates and the red star points indicate a lower shear rate) for the Newtonian and Herschel-Bulkley behavior, respectively.

For the Newtonian behavior (Figure 6A), the three PDF curves coincide, showing that $\dot{\gamma}$ has no effect on the final orientation. For the Herschel-Bulkley behavior (Figure 6B), the curves coincide for the two points of higher shear rate (rhombus and circle). However, at the point of lower shear rate (red star), the degree of orientation reduces slightly for $\theta \approx 1.7$, while the general trend is conserved, showing that only shear rates below $10^{-4}s^{-1}$ induce a slight decrease in anisotropy.

Figure 6C and 6D show a direct comparison between the two behaviors, displaying the PDF distribution at the blue rhombus and red star points, respectively, where the shear rate gradient is highest. Figure 6C shows that the constituent orientation is not affected by the fluid properties at a high shear rate. Furthermore, a preferential orientation along the flow direction $\theta \approx 1.7$ was observed for both Newtonian and Herschel-Bulkley behavior, with the same PDF mean value $\overline{\psi} = 1.66 \pm 1.31$ rad. In comparison, Figure 6D shows a slightly better alignment along the flow direction $\theta \approx 1.7$ for the Newtonian behavior at a low shear rate $\dot{\gamma}_{Ntn} = 25.90s^{-1}$, with a PDF mean value of $\overline{\psi}_{Ntn} = 1.66 \pm 1.30$ rad, compared with the Herschel-Bulkley behavior at a higher shear rate $\dot{\gamma}_{HB} = 10^{-4}s^{-1}$ with $\overline{\psi}_{HB} = 1.67 \pm 1.45$ rad. Again, those results show that only shear rates below $10^{-4}s^{-1}$ induce a slight decrease in anisotropy.

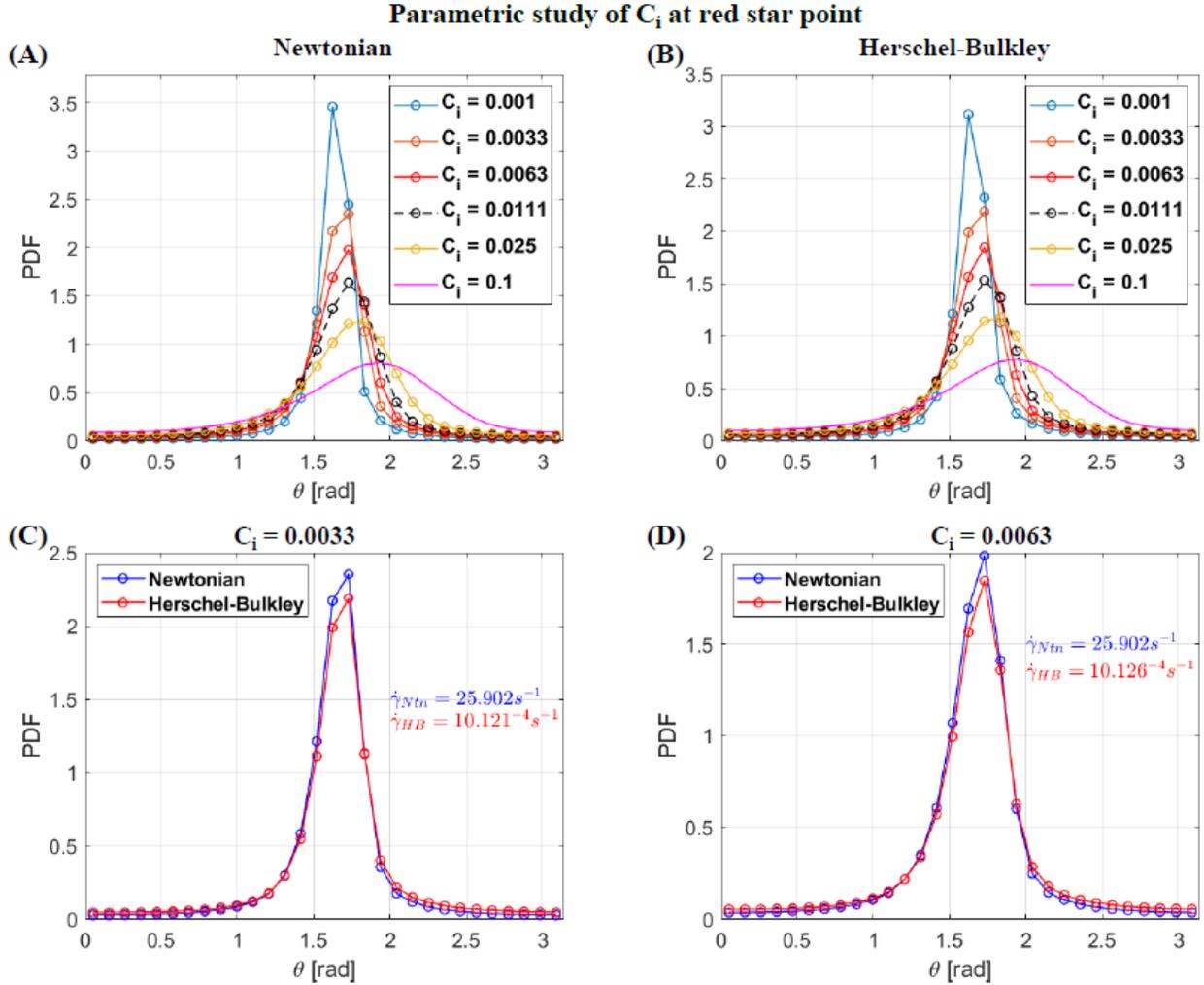

Figure 7. Parametric study of the interaction coefficient $C_i$ for the benchmark at the red star point. The study was performed for both rheologies. This figure shows the constituent PDF distribution at the red star point displayed in Figure 5. The PDF is computed with $C_i \in [0.001; 0.1]$ for the (A) Newtonian fluid and (B) Hershel Bulkley fluid. Comparison of PDF curves between Newtonian and Hershel Bulkley fluids with (C): $C_i = 0.0033$ and (D): $C_i = 0.0063$. These results are consistent with Folgar and Tucker's work [31].

In our model, the interaction between constituents can be expressed by the interaction coefficient $C_i$ in the Folgar and Tucker diffusion term equation (7), allowing us to take into account the interaction between the constituents in a non-dilute regime, such as in the viscous hydrogels used in this study. Folgar and Tucker [47] showed that the constituent orientation tends to be randomized with an increased $C_i$ value. Our results are consistent with those of Folgar and Tucker, as illustrated by the orientation of the ellipses plotted in Figure 5, which do not align perfectly along streamlines for a fixed value of $C_i = 0.01$. Therefore, we performed a parametric study of $C_i$ to assess its influence on the PDF distribution. Figure 7A and Figure 7B show that, with an increasing interaction coefficient $C_i$, the PDF distribution curve becomes wider and more dispersed, leading to a less pronounced anisotropy of constituent orientation for both rheological behaviors (Newtonian and Herschel-Bulkley). The constituent orientation $\theta$ goes from $\approx \pi/2$ to $3\pi/4$ when $C_i$ increases from 0.001 to 0.1. Notably, for small $C_i$ values, the constituents do not perfectly align with the flow direction ($\theta \neq \pi/2$) due to interaction between the constituents that cannot be totally canceled in the model due to the absence of convergence for a zero coefficient value.

To investigate the influence of the hydrogel behavior, a comparison was performed at the star point with the same given coefficient $C_i = 0.0033$ and $C_i = 0.0063$ as shown in Figure 7C and 7D, respectively. The results show that the Newtonian behavior induces a better alignment of constituents than the Herschel-Bulkley behavior. Indeed the Newtonian PDF is higher so there is less dispersion around the mean value. Specifically, as illustrated in Figure 7C with $C_i = 0.0033$, the mean PDF value is $\overline{\psi}_{Ntn} = 1.64 \pm 1.09$ and $\overline{\psi}_{HB} = 1.65 \pm 1.27$ ; and in Figure 7D, with $C_i = 0.0063$, the mean PDF value is $\overline{\psi}_{Ntn} = 1.65 \pm 1.21$ and $\overline{\psi}_{HB} = 1.66 \pm 1.37$ for the Newtonian and Herschel-Bulkley behaviors, respectively. The mean and standard deviation of the maximum PDF at the star point for the parametric study of $C_i$ in the benchmark case are displayed in Table 1 of the supplementary data.

### 3.3.2 Ink flow through a tubular nozzle syringe

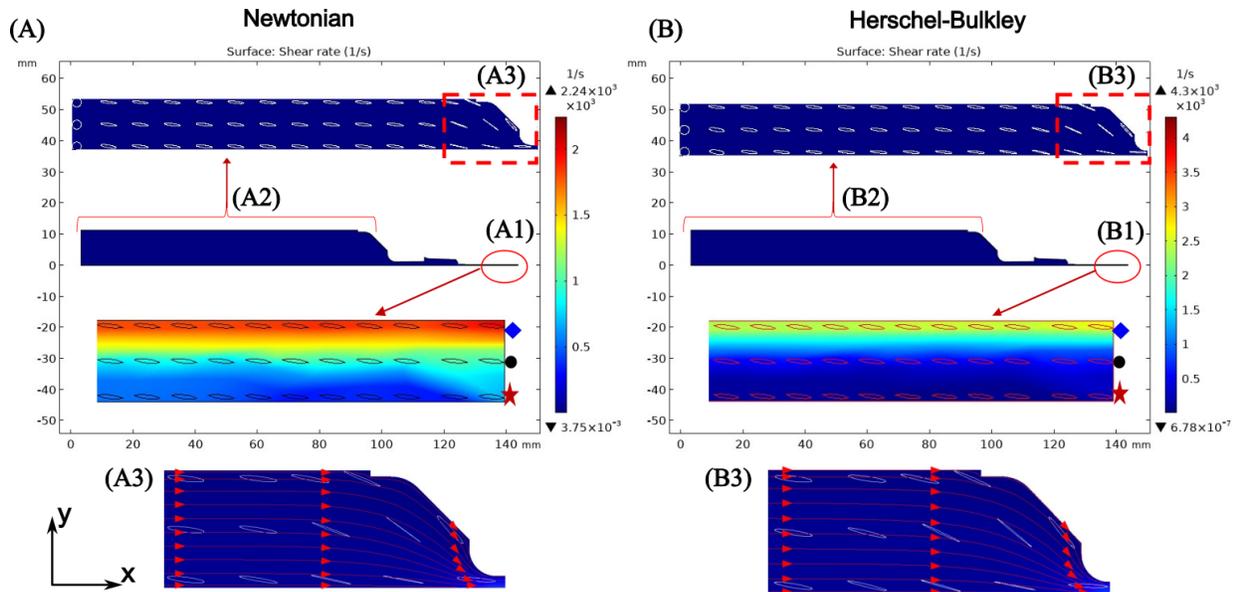

Figure 8. Shear rate distribution of fluid in a tubular nozzle syringe with interaction coefficient $C_i = 0.01$. The average constituent orientation is represented by the ellipses. (A) Newtonian behavior; (B) Hershel-Bulkley behavior. (A3) and (B3) show, in higher magnification, the rate at the constriction of the syringe. The red arrow lines indicate the streamlines of velocity.

To predict the constituent orientation in a printed scaffold, we needed to simulate an experimental-like extrusion-based 3D bioprinting process. Therefore, we tested the numerical model on a tubular nozzle syringe geometry for the two hydrogel behaviors with the extrusion parameters used in 3D bioprinting processes [3] to prevent cell death, as detailed in section 2.5.

The numerical results are shown in Figure 8, with the background color corresponding to the shear rate distribution and the ellipses in the foreground representing average constituent orientations for both Newtonian and Herschel-Bulkley rheological behaviors. Once again, the circle-like ellipses at the inlet illustrate the isotropic constituent orientation applied to the Fokker-Planck equation which corresponds to the BC1' on the left-hand side of the model applied to the Fokker-Planck equation such that $\psi = 1/\pi$. As shown in Figure 8A2 and Figure 8B2, the ellipses are then flattened and elongated, showing a preferential alignment indicated by their major axis orientation. On the right-hand side, constituents incompletely align toward the flow direction. The anisotropy is such that $\theta \neq \pi/2$ due to the interaction between the constituents modeled by the interaction coefficient $C_i = 0.01$, which is non-null for the benchmark, as discussed in the previous section. This experimental-like simulation appeared to reach steady state orientation in the A2 zone more quickly than in the benchmark, but this was only due to the length of the model. The planar channel was 4 mm long, compared with the A2 zone of the syringe, which was 100 mm long.

Figure 8A3 and 8B3 show that in the vicinity of the syringe constriction, constituents are further deviated from the streamlines depicted by the red arrow lines, compared with the left-hand side. The venturi phenomenon in this area causes a small vortex, which affects the constituent orientation leading to a deviation from the streamlines [67].

Near the convergence zone (right-hand side of A2 and B2 zones), constituents are oriented along the direction of stretch because they are close to the centerline and are subjected to elongational flow. Constituent orientation near the syringe exit is shown in the A1 and B1 zones. We observed a privileged orientation for both rheological behaviors, with constituents aligning at $\theta \approx 1.7$, as illustrated in Figure 9 by the peak at $\theta \approx 1.7$. The anisotropy is such that $\theta \neq \pi/2$ due to the interaction between the constituents modeled by the interaction coefficient $C_i = 0.01$, which is non-zero, as discussed in the previous section, and will be presented in the parametric study paragraph.

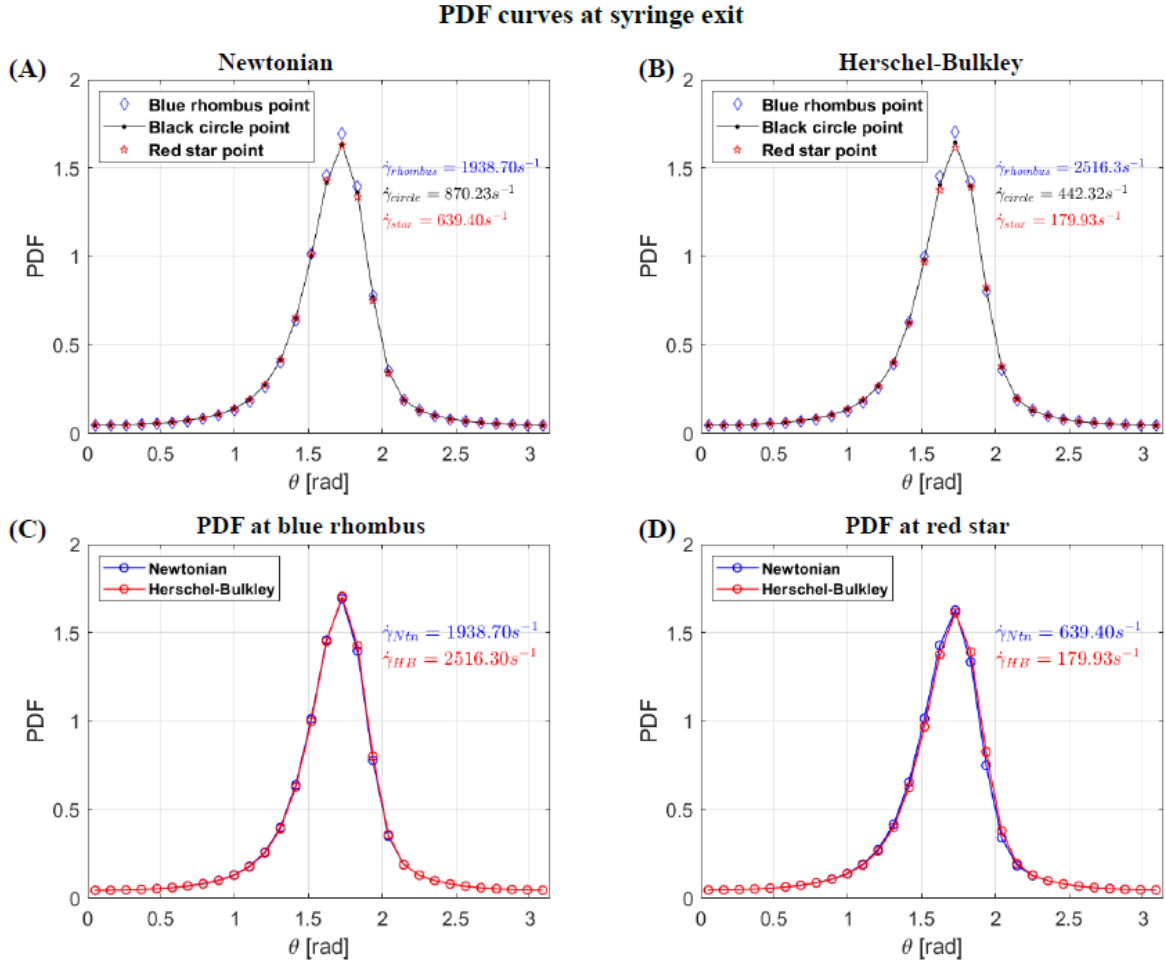

Figure 9. PDF curves for the tubular nozzle syringe case with interaction coefficient $C_i = 0.01$. Constituent orientation at three points: blue rhombus, black circle and red star in the syringe as displayed in Figure 8. (A) Newtonian fluid; (B) Herschel-Bulkley fluid. (C) and (D) PDF curves compared between Newtonian and Hershel-Bulkley behavior at blue rhombus and red star points, respectively.

PDF comparison curves at syringe exits are plotted in Figure 9. In contrast with the benchmark (Figure 6), we observed identical constituent orientation between Newtonian and Herschel-Bulkley behaviors (Figure 9A and 9B). Also, at the blue rhombus point where the shear rate is the highest, the constituents were slightly better aligned toward the flow direction, in comparison with the constituents submitted to a lower shear rate (blue rhombus *vs.* black circle *vs.* red star), due to $C_i = 0.01$ leading $\theta \neq \pi/2$. The differences between the PDF values were negligible but it is worth noting that they were in the same order of magnitude as those observed in Figure 6. However, regarding the shear rate values, there was a six order of magnitude difference in the benchmark ($\dot{\gamma}_{rhombus} = 744.28s^{-1}$ *vs.* $\dot{\gamma}_{star} = 10^{-4}s^{-1}$ as displayed in Figure 6B), whereas only a one order of magnitude difference in the syringe case ($\dot{\gamma}_{rhombus} = 2516.3s^{-1}$ *vs.* $\dot{\gamma}_{star} = 179.93s^{-1}$ as displayed in Figure 9B) for the Herschel Bulkley fluid. For the Newtonian fluid, one order of magnitude on the shear rate induced no difference in the benchmark ($\dot{\gamma}_{rhombus} = 225.47s^{-1}$ *vs.* $\dot{\gamma}_{star} = 25.90s^{-1}$ as displayed in Figure 6A), compared with less than a threefold difference in the syringe case ($\dot{\gamma}_{rhombus} = 1938.70s^{-1}$ *vs.* $\dot{\gamma}_{star} = 639.40s^{-1}$ as displayed in Figure 9A). The comparison between the Newtonian and Herschel-Bulkley fluids displayed in Figure 9C and 9D shows that the hydrogel behavior has no effect on the orientation distribution when the shear rate is in the same order of magnitude ($\dot{\gamma}_{Ntn} = 1938.7s^{-1}$ *vs.* $\dot{\gamma}_{HB} = 2516.30s^{-1}$ and $\dot{\gamma}_{Ntn} = 639.40s^{-1}$ *vs.* $\dot{\gamma}_{HB} = 179.93s^{-1}$ at the blue rhombus and the red star, respectively). There is a slight difference in the maximum PDF value at the blue rhombus (Figure 9C - $\psi_{rhombus}^{max} = 1.70$ ), compared with that at the red star (Figure 9D - $\psi_{star}^{max} = 1.61$). This can be explained by a shear rate that is lower by one order of magnitude at the red star point, compared with the blue rhombus. Nevertheless, a privileged orientation was found at the syringe exit. Most constituents aligned toward the extrusion direction at $\theta \approx 1.7$ with a mean PDF value $\overline{\psi} = 1.65 \pm 1.34$ at the blue rhombus point and $\overline{\psi} = 1.65 \pm 1.36$ at the red star point. As previously mentioned, shear rate levels were restricted to experimental-like extrusion-based 3D bioprinting process to prevent cell death.

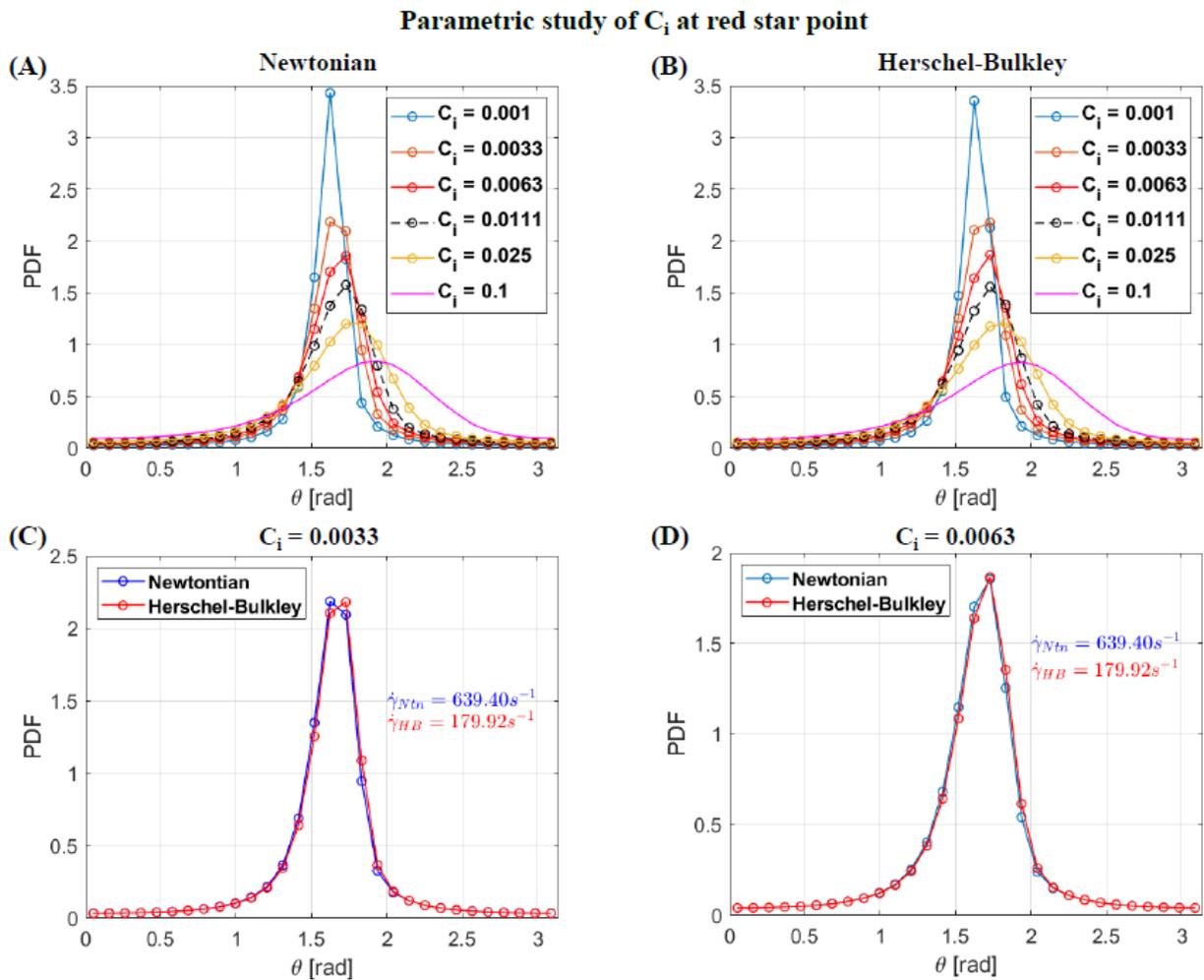

Figure 10. Parametric study of the interaction coefficient $C_i$ for the syringe case at the red star point following Folgar and Tucker's work displayed in Figure 8. PDF is computed with $C_i \in [0.001; 0.1]$ for the (A) Newtonian fluid; (B) Hershel Bulkley fluid. Comparison of PDF curves between Newtonian and Hershel Bulkley fluids with (C): $C_i = 0.0033$ and (D): $C_i = 0.0063$.

A parametric study of the interaction coefficient $C_i$ was performed at the red star point (Figure 10). As expected, constituents were more likely to be aligned with small $C_i$ values as observed in the benchmark. When $C_i$ increased, the PDF distribution curves became wider and more dispersed, leading to a less pronounced anisotropy of constituent orientation for both hydrogel behaviors (Newtonian and Herschel-Bulkley, shown in Figure 10A and 10B), similar to the benchmark results. Furthermore, with a small given coefficient $C_i = 0.0033$, although the Newtonian and Herschel-Bulkley behaviors produced the same maximum value for PDF ($\psi \approx 2.18$), there was a slight difference in mean values $\overline{\psi}_{Ntn} = 1.62 \pm 1.15$ and $\overline{\psi}_{HB} = 1.63 \pm 1.14$, respectively (Figure 10C). In contrast, when $C_i$ increased ($C_i = 0.0063$), the fluid properties no longer affected the final constituent orientation at the syringe exit and most of constituents aligned toward the extrusion direction with the same PDF mean value of $\overline{\psi} = 1.64 \pm 1.27$ (Figure 10D). Due to the empirical nature of the parameter $C_i$, there is no method to predict the $C_i$ value. Characterization of $C_i$ has been the subject of numerous investigations, which seek to establish its connection with physical properties of constituents. Notably, models such as Phan-Thien and Bay have been employed to relate $C_i$ to the aspect ratio (length and diameter) of constituents [68,69]. Adjusting this factor can achieve a targeted $C_i$ value, consequently creating the desired direction of constituents. These data show that constituent orientation is not affected by fluid properties when shear rate reaches approximately $100s^{-1}$ (Figure 9C and 9D). The interaction coefficient $C_i$ has a strong impact on constituent orientation, which overcomes the rheological behavior as illustrated in Figure 10C and Figure 10D. The mean and standard deviation of the maximum PDF at the star point for the parametric study of $C_i$ in the syringe case are displayed in Table 2 of the supplementary data.

Our data show that the interaction coefficient parameter $C_i$ plays a key role in constituent orientation and with higher $C_i$ coefficients, more constituents are dispersed from the extrusion direction. Moreover, a non-zero interaction coefficient $C_i$ prevents constituents from orienting along the fluid flow, meaning $\theta$ cannot be equal to $\pi/2$ since $C_i > 0$ for numerical convergence issues.

### 3.3.3 Comparison with experimental data

In comparing our numerical model with existing literature, Mehdipour et al. [28] and Prendergast et al. [30] provide relevant benchmarks for validation. Mehdipour et al. used a combined approach of computational fluid dynamics (CFD) to measure shear rates and Small-Angle Light Scattering (SALS) for direct observation of particle anisotropy in flow. Their work showed that shear rates around 70 s⁻¹, corresponding to a Péclet number $Pe = 20$, induced platelet orientation of $\pm10°$ relative to the flow direction. When titania rodlike fractals were introduced, under the same flow conditions, they reported a more pronounced orientation, with an average orientation angle $\chi = 1.5°$ and $Pe = 80$. In our study, the Péclet numbers ($Pe$) obtained from the simulation ranged between [9.6 to 120], placing them within the same range as the Mehdipour study. Our numerical results produced a mean PDFvalue $\overline{\psi}$ centered in $\theta \cong 1.7$ rad (~7° along the direction of the flow), closely matches their results. This slight discrepancy can be attributed to differences in viscosity: their study involved viscosities ranging from 0.01 to 0.1 Pa.s, while in our case, the hydrogel viscosity is significantly higher (0.8 Pa.s in the channel geometry and 10 Pa.s in the syringe geometry), which dampens alignment.

In a related context, Prendergast et al. [30] employed a numerical approach using COMSOL Multiphysics to predict flow rates and shear stress profiles for different bioink formulations, optimizing printing parameters. Their study demonstrated a clear correlation between the shear stress profiles and fiber alignment observed in printed collagen samples. By using scanning electron microscopy (SEM), they confirmed that collagen fibers aligned along the flow direction, particularly under conditions of higher inlet flow rates, which increased fiber concentrations in the bioink. These results validate our own findings, where the predicted relationship between shear stress and constituent alignment was corroborated by the mean PDF results, showing constituent alignment along the flow direction. Our approach, however, goes further by establishing a direct relationship between shear rate and constituent orientation, offering a predictive model for fibrillar hydrogel anisotropy, which was not captured through numerical simulation in their study.

While both Kim et al. [32] and Nerger et al. [34] used experimental approaches to quantify shear rates, their work further supports the findings of Mehdipour and Prendergast. Kim and Nerger utilized analytical models to estimate shear stress and flow rates, validating the direct relationship between shear flow and fiber alignment. Although not based on numerical modeling, these studies provide experimental evidence that aligns well with our computational predictions, thereby reinforcing the reliability of our model in predicting fibrillar hydrogel anisotropy.

In summary, our numerical model aligns well with experimental observations from the literature, demonstrating its potential as a robust tool for predicting anisotropy in hydrogels. Unlike previous studies that only focus on fluid-flow modeling applied to hydrogels, our approach directly links shear rate to constituent orientation, providing a unique in silico method for quantifying the anisotropic behavior of fibrillar hydrogels under flow. This capability could be valuable for optimizing 3D bioprinting processes and enhancing control over scaffold properties.

### 3.4 Limitations and future directions

In this study, we developed the first numerical model able to compute and predict flow-induced orientation within a hydrogel in the field of bioprinting. Therefore, the computational model was restricted to steady flow and orientation in 2D to simplify the analysis. Most of the biological hydrogels, composed of different molecules, may exhibit viscoelastic behavior [70] as well as shear thinning effects. To reduce the complexity of the numerical model, and the number of parameters involved, this viscoelastic behavior is neglected.

The presented framework should be viewed as a first approach, offering insights into anisotropy while balancing computational constraints through the adoption of a one-way coupling approach for the flow and polymer orientation, a choice driven by the need to balance model accuracy and computational feasibility. While fully coupled models can capture more detailed fluid-particle interactions, studies such as Mezi et al. [71] and Kermani et al. [72] have demonstrated that, the differences in polymer orientation between coupled and uncoupled simulations are minimal. For example, Mezi et al. [71] showed that the coupling coefficient does not significantly influence the final fiber orientation, while Wang et al. [73] found that the fiber alignment in the flow direction increases by no more than 17% in fully coupled simulations. These results indicate that one-way coupling remains a valid approach for predicting fiber alignment in extrusion-based systems, particularly for viscous fluids in semi-dilute regimes and low-shear-rate regimes common in bioprinting.

To model interactions in semi-concentrated solutions, we introduced the interaction coefficient ($C_i$), following the approach of Folgar and Tucker. This diffusion term accounts for particle interactions, extending the Jeffery model beyond the dilute regime. Previous studies, such as those by Phan-Thien et al. [68], Heller et al. [74] and Berin Seta et al. [75], have validated this approach, showing that the Jeffery equation, combined with the $C_i$ coefficient, provides a good approximation for fiber rotation rates in semi-dilute systems. While we acknowledge that our assumptions simplify the modeling of concentrated solutions, the combination of one-way coupling and the $C_i$ coefficient has been widely used in the literature as a reasonable initial framework for studying fiber orientation under flow. This provides a computationally efficient method to analyze anisotropy in hydrogel scaffolds without sacrificing essential predictive capabilities.

Additionally, the simplified rigid rod (or dumbbell) model used to represent the polymer chains neglects important physical effects such as chain stretching and entanglement, which become relevant at higher concentrations. While this simplification is not fully representative of real polymer dynamics, as highlighted by Doi and Edwards, it provides a useful framework for understanding how macromolecular motion influences rheological behavior in a tractable manner. More complex models that account for polymer elasticity and entanglement would offer a more accurate representation of polymer behavior, but they require a significant computational investment. Future iterations of this work will focus on incorporating these effects, as well as the full two-way coupling, to extend the applicability of the model to a wider range of concentrations and better reflect realistic polymer dynamics in 3D bioprinting scenarios. Additionally, while the interaction coefficient $C_i$ has been demonstrated to significantly impact constituent orientation, this study focused on a fixed $C_i$ value as a parameter in the model.

The outcomes from this study will allow us to explore two main perspectives in future work. First, an experimental analysis of X-ray synchrotron data is underway to quantify the constituent orientation by Herman's factor calculation [76,77] and validate the numerical model prediction. Second, as an important factor influencing the orientation model, many investigations have been carried out to express the physical meaning of the interaction coefficient $C_i$ from experimental observations. Therefore, it will be necessary to run the simulation with $C_i$ computed from a different model to discriminate the physical properties of the hydrogel that best describe the interaction between constituents.

## 4. Conclusion

In this study, we developed a numerical strategy to predict constituent orientation in a hydrogel during an extrusion process by directly solving coupled problems of fluid flow and constituent motion. This dual capability sets our model apart from previous studies which focused primarily on shear stress or empirical observations without offering a predictive model for fibrillar hydrogel anisotropy. Our approach bridges this gap by directly linking shear rate to constituent alignment, providing a more comprehensive framework for predicting anisotropy in bioprinted hydrogels.

One of the key findings of our model is that high shear rates dominate the rheological behavior of the hydrogel, driving constituent orientation along the flow direction. This result underscores the importance of shear rate in the final alignment of hydrogel fibers, even as fluid rheological properties play a secondary role under these conditions. Additionally, our study highlights the critical role of the interaction coefficient $C_i$, which represents microscopic interactions between fluid particles [47]. As $C_i$ increases, the orientation distribution curves become wider and more dispersed, leading to a less pronounced anisotropy in constituent orientation. This effect is essential for understanding how microscopic interactions influence the macroscopic structure of bioprinted scaffolds.

Moreover, our data suggests that, at the low shear rates typically found during extrusion-based 3D bioprinting, the shear rate distribution and hydrogel behavior have almost no influence on the final constituent orientation at the syringe exit. Most of the constituents were aligned toward the direction of extrusion.

# Acknowledgments

The authors thank Leah Cannon for the English language editing of this manuscript.

# Declaration of interest statement

The authors declared no potential conflicts of interest with respect to the research, authorship, and/or publication of this article.

**Funding**

**CRediT authorship contribution statement**

Van Than Mai: Conceptualization; Data curation; Formal analysis; Investigation; Methodology; Software; Validation; Visualization; Roles/Writing - original draft; Writing - review & editing.

Robin Chatelin: Methodology; Software; Roles/Writing - original draft; Writing - review & editing,

Edwin-Joffrey Courtial: Conceptualization; Resources; Writing - review & editing.

Caroline Boulocher: Conceptualization, methodology, validation, writing - review & editing, supervision, project administration, funding acquisition.

Romain Rieger: Conceptualization; Formal analysis; Investigation; Methodology; Project, supervision, Administration; Software; Validation; Visualization; Roles/Writing - original draft; Writing - review & editing.

**ORCID**

Van Than MAI: 0000-0001-8579-1670

Robin Chatelin: 0000-0001-6504-941X

Edwin-Joffrey Courtial: 0000-0002-1406-5871

Caroline Boulocher:  0000-0002-7202-420X

Romain Rieger: 0000-0003-2792-1451

**Glossary**

BC: boundary conditions

E: east

FAG: fibrinogen alginate gelatin

FEM: finite element method

FENE: finitely extensible non-linear elastic

FVM: finite volume method

HB: Hershel-Bulkley

Ntn: Newtonian

PMMA: poly(methyl methacrylate)

PLLA: poly(L-lactic acid)

PDF: probability distribution function

Re: Reynolds number

REV: representative elementary volume

TE: tissue engineering

W: west

Hydrogel Constructs by 3D Printing for Application in the Engineering of Mechanically Demanding Tissues, Polymers 13 (2021) 1663. https://doi.org/10.3390/polym13101663.

## Appendix: Upwind differencing and Power law schemes

The upwind differencing and power law schemes allowed us to calculate the coefficients of the discretization form of Fokker-Planck equation 9 as follows:

$$a_P = a_E + a_W + F_e - F_w$$

$$a_E = D_e \, max\left[0, (1-0.1|\frac{F_e}{D_e}|)^5\right] + max[-F_e, 0]$$

$$a_W = D_w \, max\left[0, (1-0.1|\frac{F_w}{D_w}|)^5\right] + max[F_w, 0]$$

Where $F_e$ and $F_w$ represent the motion of a constituent at the east and west side control volume face, respectively. The $F_e$ and $F_w$ are determined from the Jeffery equation 6 following as:

$$F_e = (cos\theta \quad -sin\theta)_e \begin{pmatrix} 0 & W_{12} \\ -W_{12} & 0 \end{pmatrix}_e \begin{pmatrix} sin\theta \\ cos\theta \end{pmatrix}_e + \lambda(cos\theta \quad -sin\theta)_e \begin{pmatrix} D_{11} & D_{12} \\ D_{12} & -D_{11} \end{pmatrix}_e \begin{pmatrix} sin\theta \\ cos\theta \end{pmatrix}_e$$

$$F_w = (cos\theta \quad -sin\theta)_w \begin{pmatrix} 0 & W_{12} \\ -W_{12} & 0 \end{pmatrix}_w \begin{pmatrix} sin\theta \\ cos\theta \end{pmatrix}_w + \lambda(cos\theta \quad -sin\theta)_w \begin{pmatrix} D_{11} & D_{12} \\ D_{12} & -D_{11} \end{pmatrix}_w \begin{pmatrix} sin\theta \\ cos\theta \end{pmatrix}_w$$

and $D_e = D_w = D_r/\delta\theta$

## Supplementary data

|  | $C_i = 0.001$ | $C_i = 0.0033$ | $C_i = 0.0063$ | $C_i = 0.0111$ | $C_i = 0.025$ | $C_i = 0.1$ |
|---|---|---|---|---|---|---|
| $\overline{\psi}_{Ntn}$ | $1.63 \pm 0.88$ | $1.64 \pm 1.09$ | $1.65 \pm 1.21$ | $1.67 \pm 1.33$ | $1.69 \pm 1.5$ | $1.76 \pm 1.88$ |
| $\overline{\psi}_{HB}$ | $1.64 \pm 1.12$ | $1.65 \pm 1.27$ | $1.66 \pm 1.37$ | $1.67 \pm 1.47$ | $1.69 \pm 1.63$ | $1.76 \pm 1.95$ |

Table 1. Mean and standard deviation of the maximum PDF at the star point for parametric study of $C_i$ in the benchmark. The results show that when $C_i$ is fixed, the mean PDF values are similar for both rheological behaviors. However, the Herschel-Bulkley behavior exhibits a larger standard deviation compared to the Newtonian behavior. The fact that the standard deviation is larger for Herschel-Bulkley means that the PDF values in this behavior are more dispersed around the mean, leading to reduced anisotropy of the constituents.

|  | $C_i = 0.001$ | $C_i = 0.0033$ | $C_i = 0.0063$ | $C_i = 0.0111$ | $C_i = 0.025$ | $C_i = 0.1$ |
|---|---|---|---|---|---|---|
| $\overline{\psi}_{Ntn}$ | $1.61 \pm 0.97$ | $1.62 \pm 1.15$ | $1.64 \pm 1.27$ | $1.65 \pm 1.38$ | $1.68 \pm 1.56$ | $1.75 \pm 1.94$ |
| $\overline{\psi}_{HB}$ | $1.62 \pm 0.95$ | $1.63 \pm 1.14$ | $1.64 \pm 1.27$ | $1.66 \pm 1.38$ | $1.68 \pm 1.55$ | $1.76 \pm 1.89$ |

Table 2. Mean and standard deviation of the maximum PDF at the star point for parametric study of $C_i$ in the syringe case. The results show that when Ci is fixed, unlike in the benchmark, the mean PDF values and the standard deviation are similar for all rheological behaviors. These results suggest that for the syringe with a tubular nozzle, the orientation of the constituents is not affected by the rheological behaviors of the hydrogel.